\documentclass{emulateapj}
\usepackage{verbatim} 

\shorttitle{The redshift and nature of AzTEC/COSMOS~1}
\shortauthors{V. Smol\v{c}i\'{c} et al.}

\def\f#1   {Fig.~\ref{#1}}
\def\s#1   {Sec.~\ref{#1}}
\def\tab#1   {Tab.~\ref{#1}}
\def\t#1   {Tab.~\ref{#1}}

\def\comm#1   {{\tt (COMMENT: #1) }}

\def\kms{~km~s$^{\mathrm{-1}}$}

\def\i                {{\em i}}

\def\asec              {$''$}

\def\lsol              {$\mathrm{L}_{\odot}$}

\def\msol              {$\mathrm{M}_{\odot}$}

\def\msolyr              {$\mathrm{M}_{\odot}\, \mathrm{yr}^{-1}$}
\def\msolyrkpc              {$\mathrm{M}_{\odot}\, \mathrm{yr}^{-1}\, \mathrm{kpc}^{-2}$}

\def\kms{~km~s$^{\mathrm{-1}}$}

\def\smo                   {Smol\v{c}i\'{c}}

\slugcomment{  }

\begin{document}

\title{ The redshift and nature of AzTEC/COSMOS~1: A starburst galaxy at $z=4.6$}

\author{
  V.~Smol\v{c}i\'{c}\altaffilmark{1,2},
  P.~Capak\altaffilmark{3,4}, 
  O.~Ilbert\altaffilmark{5}, 
  A.W.~Blain\altaffilmark{3},
  M. Salvato\altaffilmark{3,9},
I.~Aretxaga\altaffilmark{14}, 
 E.~Schinnerer\altaffilmark{23}, 
  D.~Masters\altaffilmark{3,6},
  I.~Mori\'{c}\altaffilmark{3,7},
  D.~A.~Riechers\altaffilmark{3,8},
 K.~Sheth\altaffilmark{4},
 M.~Aravena\altaffilmark{10},
 H.~Aussel\altaffilmark{11},
 J.~Aguirre\altaffilmark{12,13},
S.~Berta\altaffilmark{15},
 C.~L.~Carilli\altaffilmark{16},
 F.~Civano\altaffilmark{17},
 G.~Fazio\altaffilmark{17},
 J.~Huang\altaffilmark{17},
 D.~Hughes\altaffilmark{14},
 J.~Kartaltepe\altaffilmark{18},
 A.~M.~Koekemoer\altaffilmark{19},
  J.-P. Kneib\altaffilmark{5},
  E.~LeFloc'h\altaffilmark{18,20},
D.~Lutz\altaffilmark{15},
 H.~McCracken\altaffilmark{5}, 
   B.~Mobasher \altaffilmark{6},
 E.~Murphy\altaffilmark{4}, 
F.~Pozzi\altaffilmark{21},
L.~Riguccini\altaffilmark{22},
 D.~B.~Sanders\altaffilmark{18},
M.~Sargent\altaffilmark{23},
 K.~S.~Scott\altaffilmark{24},
 N.Z.~Scoville\altaffilmark{3},
 Y.~Taniguchi\altaffilmark{25},
 D.~Thompson\altaffilmark{26},
 C.~Willott\altaffilmark{27},
 G.~Wilson\altaffilmark{28},
 M.~Yun\altaffilmark{28}
    }

\altaffiltext{1}{ESO ALMA COFUND Fellow, European Southern Observatory, Karl-Schwarzschild-Strasse 2, 
85748 Garching b. Muenchen, Germany}
\altaffiltext{2}{Argelander Institut for Astronomy, Auf dem H\"{u}gel 71, Bonn, 53121, Germany}
\altaffiltext{3}{ California Institute of Technology, MC 249-17, 1200 East
California Boulevard, Pasadena, CA 91125 }
\altaffiltext{4}{Spitzer Science Center, 314-6 Caltech, 1201 E. California Blvd. Pasadena, CA, 91125}
\altaffiltext{5}{Laboratoire d'Astrophysique de Marseille, Universit\'e de
Provence, CNRS, BP 8, Traverse du Siphon, 13376 Marseille Cedex 12, France}
\altaffiltext{6}{Department of Physics and Astronomy, University of California, Riverside, CA, 92521, USA}
\altaffiltext{7}{University of Zagreb, Physics Department, Bijeni\v{c}ka cesta 32, 10000 Zagreb, Croatia}
\altaffiltext{8}{Hubble Fellow}
\altaffiltext{9}{Max-Planck-Institut fÃ¼r Plasmaphysik, Boltzmanstrasse 2,
Garching 85748, Germany}
\altaffiltext{10}{National Radio Astronomy Observatory, 520 Edgemont Road, Charlottesville, VA 22903, USA}
\altaffiltext{11}{UMR AIM (CEA-UP7-CNRS), CEA-Saclay, Orme des Merisiers, bÃ¢t. 709, F-91191 Gif-sur-Yvette Cedex, France }
\altaffiltext{12}{Jansky Fellow, National Radio Astronomy Observatory}
\altaffiltext{13}{University of Pennsylvania, Department of Physics and Astronomy, 209 South 33rd Street, Philadelphia, PA 19104}
\altaffiltext{14}{Instituto Nacional de Astrof\'{\i}sica, \'Optica y Electr\'onica (INAOE), Aptdo. Postal 51 y 216, 72000 Puebla, Pue., Mexico}
\altaffiltext{15}{Max-Planck-Institut f\"{u}r extraterrestrische Physik, Postfach 1312, 85741 Garching, Germany}
\altaffiltext{16}{National Radio Astronomy Observatory, P.O. Box 0, Socorro,
  NM 87801-0387 } 
\altaffiltext{17}{Harvard-Smithsonian Centre for Astrophysics, 60 Garden Street, Cambridge, MA 02138, USA}
\altaffiltext{18}{Institute for Astronomy, University of Hawaii, 2680 Woodlawn Drive, Honolulu, HI, 96822, USA}
\altaffiltext{19}{Space Telescope Science Institute, 3700 San Martin Drive, Baltimore, MD 21218}
\altaffiltext{20}{Spitzer Fellow}
\altaffiltext{21}{INAF – Osservatorio Astronomico di Roma, via di Franscati 33, 00040 Monte Porzio Catone, Italy}
\altaffiltext{22}{Laboratoire AIM-Paris-Saclay, CEA/DSM/Irfu – CNRS – Universite Paris Diderot, CE-Saclay, pt courrier 131, F-91191 Gif-sur-Yvette, France}
\altaffiltext{23}{ Max Planck Institut f\"ur Astronomie, K\"onigstuhl 17,
  Heidelberg, D-69117, Germany } 
\altaffiltext{24}{Department of Physics and Astronomy, University of Pennsylvania,Philadelphia PA, 19104}
\altaffiltext{25}{Research Center for Space and Cosmic Evolution, Ehime University, Bunkyo-cho 2-5, Matsuyama 790-8577, Japan}
\altaffiltext{26}{Large Binocular Telescope Observatory, University of Arizona, 933 N. Cherry Ave., Tucson, AZ, 85721, USA}
\altaffiltext{27}{Herzberg Institute of Astrophysics, National
Research Council, 5071 West Saanich Rd., Victoria, BC V9E 2E7, Canada}
\altaffiltext{28}{ Department of Astronomy, University of Massachusetts, Amherst, MA 01003, USA}
\altaffiltext{$\star$}{Based on observations with:  the W.M. Keck Observatory,  the Canada-France-Hawaii Telescope; the United Kingdom Infrared Telescope; the Subaru Telescope; the NASA/ESA {\em Hubble Space Telescope}; the NASA Spitzer Telescope; the Caltech Sub-mm Observatory; the Smithsonian Millimeter Array; and the National Radio Astronomy Observatory. Herschel is an ESA space observatory with science instruments provided by European-led Principal Investigator consortia and with im- portant participation from NASA.}

\begin{abstract}
  Based on broad/narrow-band photometry and Keck DEIMOS
    spectroscopy we report a redshift of $z=4.64_{-0.08}^{+0.06}$ for
  AzTEC/COSMOS~1, the brightest sub-mm galaxy in the
  AzTEC/COSMOS field.  In addition to the COSMOS-survey X-ray to radio
  data, we report observations of the source with Herschel/PACS (100,
  160~$\mu$m), CSO/SHARC~II (350~$\mu$m), CARMA and PdBI (3~mm). We do
  not detect CO($5\rightarrow4$) line emission in the covered redshift
  ranges, 4.56-4.76 (PdBI/CARMA) and 4.94-5.02 (CARMA). If the line is
  within this bandwidth, this sets $3\sigma$ upper limits on the gas
  mass to $\lesssim8\times10^9$~\msol\ and
  $\lesssim5\times10^{10}$~\msol , respectively (assuming similar
  conditions as observed in $z\sim2$ SMGs). This could be explained by
  a low CO-excitation in the source. Our analysis of the UV-IR
  spectral energy distribution of AzTEC~1 shows that it is an
  extremely young ($\lesssim50$~Myr), massive
  ($M_*\sim10^{11}$~\msol), but compact ($\lesssim2$~kpc) galaxy
  forming stars at a rate of $\sim1300$~\msolyr . Our results imply
  that AzTEC~1 is forming stars in a 'gravitationally bound' regime in
  which gravity prohibits the formation of a superwind, leading to
  matter accumulation within the galaxy and further generations of
  star formation.
\end{abstract}


\section {Introduction}
\label{sec:introduction}

Submillimeter galaxies (SMGs; $S_\mathrm{850\mu m} > 5$~mJy) are
ultra-luminous, dusty starbursting systems with extreme star formation
rates ($\mathrm{SFR}\sim100-1000$~\msolyr ;
e.g.~\citealt{blain02}). The bulk of this population has been shown to
lie at $2<z<3$ \citep[e.g.][]{chapman05}. However, only recently have
blank-field sub-mm surveys started to discover the high-redshift
($z>4$) tail of the SMG distribution. To date seven $z>4$ SMGs
  have been spectroscopically confirmed (and published: three in
  GOODS-N, \citealt{daddi09a,daddi09b}; two in COSMOS,
  \citealt{capak08,schinnerer08,riechers10,capak10}, one in ECDFS,
  \citealt{coppin09,coppin10}; and one in Abell~2218,
  \citealt{knudsen10}). These high-redshift SMGs, presenting a
challenge to cosmological models of structure growth \citep[see
e.g.][]{coppin09}, may alter our understanding of the role of SMGs in
galaxy evolution.

Galaxies are thought to evolve in time from an initial stage with
irregular/spiral morphology towards passive, very massive elliptical
systems ($M_*>10^{11}$~\msol ; \citealt[e.g.][]{faber07}). The
morphology and spectral properties of passive galaxies indicate that
they have formed in a single intense burst at $z > 4$
\citep[e.g.][]{cimatti08}. SMGs represent short-lasting ($<< 100$~Myr)
starburst episodes of the highest known intensity. Thus, they would be the perfect candidates for
$z\sim2$ passive galaxy progenitors. In this Letter we report on a new
$z>4$ SMG -- AzTEC/COSMOS~1 (AzTEC~1 hereafter), the brightest SMG
detected in the AzTEC-COSMOS field (Scott et al.\ 2008).

We adopt $H_0=70,\, \Omega_M=0.3, \Omega_\Lambda = 0.7$, use a
Salpeter initial mass function, and AB magnitudes.

\section {Data}
\label{sec:data}

The available photometric (X-ray--radio) data for AzTEC~1
($\alpha=09:59:42.863$, $\delta=+02:29:38.19$) are summarized in
\t{tab:phot} . Its optical/IR counterpart -- identified by
\citet{younger07} in follow-up SMA observations of the original
JCMT/AzTEC 1.1~mm detection \citep{scott08} -- has been targeted by
the COSMOS project \citep{scoville07} in more than 30
filters:. ground-based optical/NIR imaging in 22 bands
\citep{capak07}\footnote{An updated version of the UV-NIR catalog,
  available at
  http://irsa.ipac.caltech.edu/data/COSMOS/tables/photometry, has been
  used.}, Chandra \citep{elvis09}, GALEX \citep{zamojski07}, HST
\citep{scoville07, koekemoer07, leauthaud07}, Spitzer
\citep{sanders07}, and VLA \citep[\smo\ et
al.,~in~prep.]{schinnerer07,schinnerer10} (see \t{tab:phot} ).

Herschel (100~and~160~$\mu$m) data are drawn from the PACS
Evolutionary Probe observations (Lutz et al.,~in~prep; 
Berta~et~al.~2010).

Observations at $350~\mu$m with CSO/SHARC~II were obtained during
  two nights in March/2009 with an average 225~GHz opacity of 
  $\tau_{225}<0.05$.  The data were reduced using the standard CRUSH
tool.  A total of $\sim6$~hrs of integration time reached an rms of 10~mJy. Combined with previous data (Aguirre et
  al., in prep) we detect no flux at $1\sigma = 7$~mJy.

Observations at 3~mm were obtained with CARMA in E-array configuration
in July/2009. The target was observed for 8.5~hrs
on-source. The 3~mm receivers were tuned to 98.95~GHz (3.03~mm), with
lower~(upper) sidebands centered at 96.43~(101.46)~GHz,
respectively. Each sideband was observed with 45 31.25\,MHz wide
channels, leading to a total bandwidth of
2.56\,GHz. The data reduction was performed with the MIRIAD
package. No line emission (the CO(5$\rightarrow$4) transition is
  expected at the source's redshift) was detected across the
  observed bands covering $4.64<z<4.72$ and $4.94<z<5.02$. The
uv-data were imaged merging both sidebands together and using natural
weighting. We infer an rms of 0.36~mJy/beam in the continuum map, but
no detection of the source.

Using the new WideX correlator on PdBI, AzTEC-1 was observed with 6
antennas in Apr/May/2010 for $\sim5.5$~hrs on-source. The
WideX correlator covered 3.6~GHz bandwidth using
polarizations centered at 101.866394 GHz. 1005+066 and 3C273 were used
as phase and gain
calibrators, respectively. The flux calibration error is estimated to
be $<10\%$.  The naturally weighted beam is 6.38''$\times$5.01''~(PA
32$^o$).  The 3~mm continuum emission, shown in \f{fig:pdbi} , is
detected at 7.5$\sigma$ with $\rm S_{3mm}=0.3\pm 0.04\,mJy$ and
unresolved.  No line emission is detected across the 
band covering $\rm 4.56 < z < 4.76 $. The rms per 180\,km/s wide
channel (61.2 MHz) is 0.35\,mJy/beam.

\begin{figure}[ht!]
\center{\vspace{8mm}
\includegraphics[width=\columnwidth, angle=-90, scale=0.9]{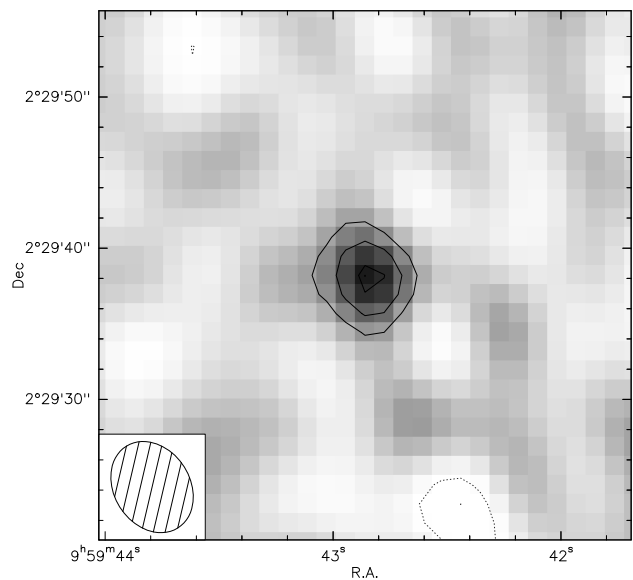}
\caption{PdBI 3~mm continuum image of AzTEC~1. Contours are at
  $\pm3\sigma$, $\pm5\sigma$, $\pm7\sigma$ (1$\sigma =
  0.04$~mJy/beam). The inset shows the clean beam. 
  \label{fig:pdbi}}
}
\end{figure}

AzTEC~1 was spectroscopically targeted with DEIMOS on
Keck-II in Nov/2008 with
clear conditions and $\sim1$\asec\ seeing and a 4~hr 
integration time split into 30~min exposures. The data were collected with
the 830l/mm grating tilted to 7900~\AA\ and the OG550 blocker.  The
objects were dithered $\pm3$\asec along the slit to remove ghosting.

The data were reduced via the modified DEEP2 DEIMOS
pipeline (see~\citealt{capak08}).  The overall
instrumental throughput was determined using the standard stars HZ-44
and GD-71.  Bright stars in the mask were used to determine the amount
of atmospheric extinction, wavelength dependent slit losses from
atmospheric dispersion, and to correct for the A, B, and water absorption
bands.  The 2D- and 1D- spectra are shown in \f{fig:spec} . No strong
emission lines are present in the spectrum. The continuum is clearly
detected (see 2D-spectrum in \f{fig:spec} ), however at low signal-to-noise,
consistent with the faint magnitude of the source ($i^+=25.2$).

\section{The redshift of AzTEC~1}
\label{sec:redsft}

From features in the DEIMOS spectrum we determine a redshift for
AzTEC~1 of $4.650\pm0.005$ based on the blue cut-off of Ly$\alpha$.
Note that in this redshift range, Ly$\alpha$, the most prominent
emission line that may be expected, would be attenuated by the atmospheric
B-band (6860-6890~\AA). The 1216~\AA\ Ly$\alpha$ forest break is
however clearly seen in the 2D-spectrum, as well as in the heavily
smoothed 1D-spectrum (see \f{fig:spec} , \f{fig:specbinn} ).  If this
were the 4000~\AA\ break at $z=0.71$ we would expect strong $160~\mu$m
and $350~\mu$m detections for any known galaxy type. As these do not
exist for AzTEC~1, low redshifts ($z<1$) can be ruled out. Note that
the inferred high redshift is consistent with both, the source being a
B-band drop-out, and its FIR/radio ratio (Younger et al.\ 2008; Yun \&
Carilli 2002).

Due to a) the low S/N, b) the general absence of strong emission
lines, and c) the atmospheric B-band bias at the expected position of
Ly$\alpha$, we utilize the photometric data available for AzTEC~1
along with the spectrum to refine our redshift estimate.  Using 31
NUV-NIR photometric measurements (Tab.~1) and the binned spectrum we
constrain the redshift via a $\chi^2$ minimization SED fitting
technique described in detail by Ilbert et al.\ (2009). Our best fit
results, as well as the redshift probability [$\exp(-\chi^2/2)$]
distribution, are shown in \f{fig:specbinn} . We find a redshift of
$z=4.64_{-0.08}^{+0.06}$, where the errors are drawn from the 68\%
confidence interval. Note that this analysis yields also a secondary
redshift peak at $z=4.44$, albeit with a significantly lower
probability than that at $z=4.64$

As it is possible that heavy extinction in the UV biases
  UV-NIR-derived photometric redshifts towards higher values, we
  estimate the photometric redshift using FIR-radio data via
  a Monte-Carlo approach, described in detail in Aretxaga et al.\
  (2003). We find that the upper limits at $\lambda <450~\mu$m
  strongly suggest $z>4.0$ (at $\sim~90\%$ confidence). The
  redshift probability distribution reaches a plateau with equally
  plausible solutions between $z=4.5$ and $z=6.0$, supporting the
  optical-IR redshift solution.  

The inferred most probable redshift $z=4.64$ (based on UV-NIR
data) is close to the spectroscopically determined redshift of
$z=4.65$, and supported by the FIR-radio data.  Thus, hereafter we
take $z=4.64_{-0.08}^{+0.06}$ as the best estimate for the redshift of
AzTEC~1.

\begin{figure*}[ht!]
\center{\vspace{8mm}
\includegraphics[bb= 0 0 1178 233, scale=0.4]{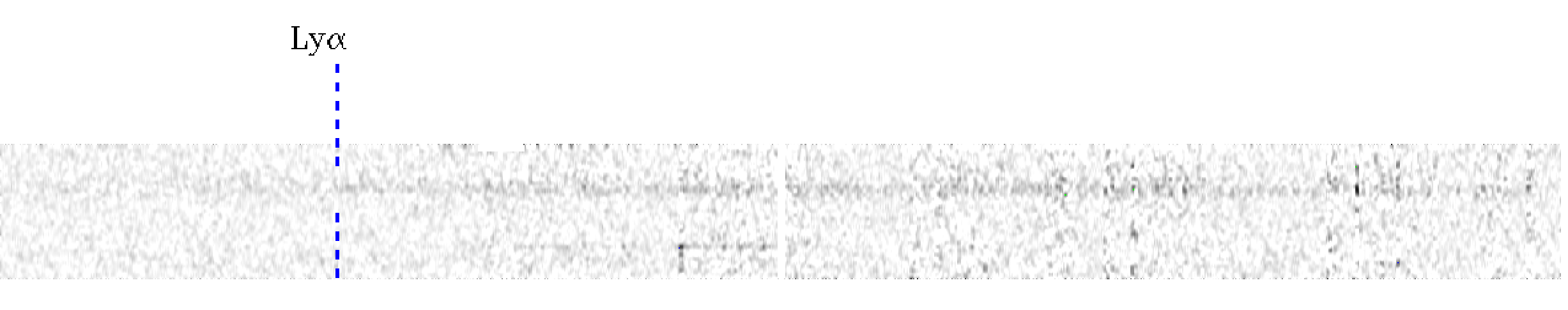}\\
\vspace{-5mm}
\includegraphics[scale=0.6]{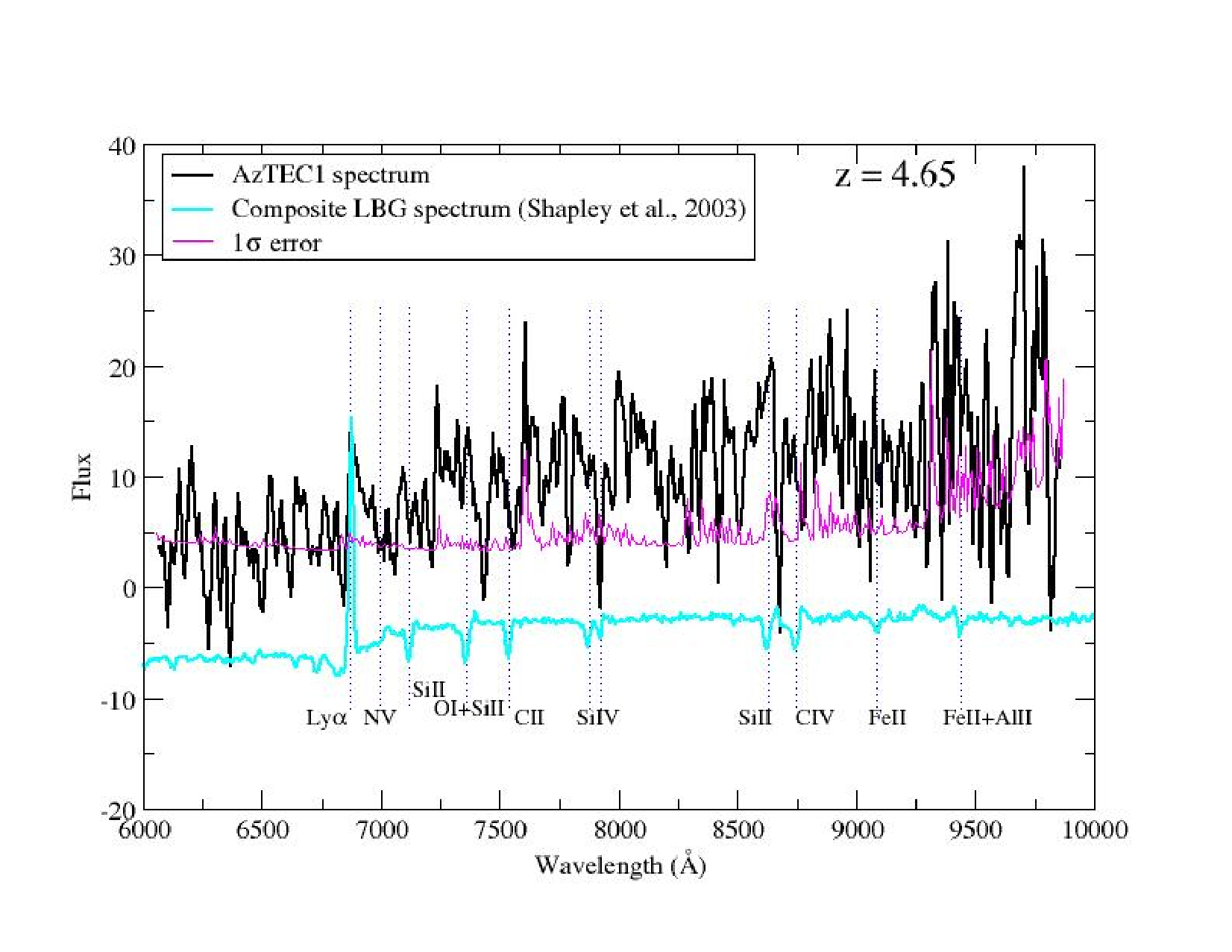}
\caption{The top panel shows the Keck~II/DEIMOS 2D-spectrum of
  AzTEC~1. Note the increase in continuum flux beyond Ly$\alpha$
  (see also \f{fig:specbinn} ). In the bottom panel the extracted 1D-spectrum is shown. Note that the atmospheric B-band
    (6860-6890~\AA) is coincident with the expected Ly$\alpha$ emission line.
  \label{fig:spec}}
}
\end{figure*}

\begin{figure}[ht!]
\center{\vspace{8mm}
\includegraphics[width=\columnwidth]{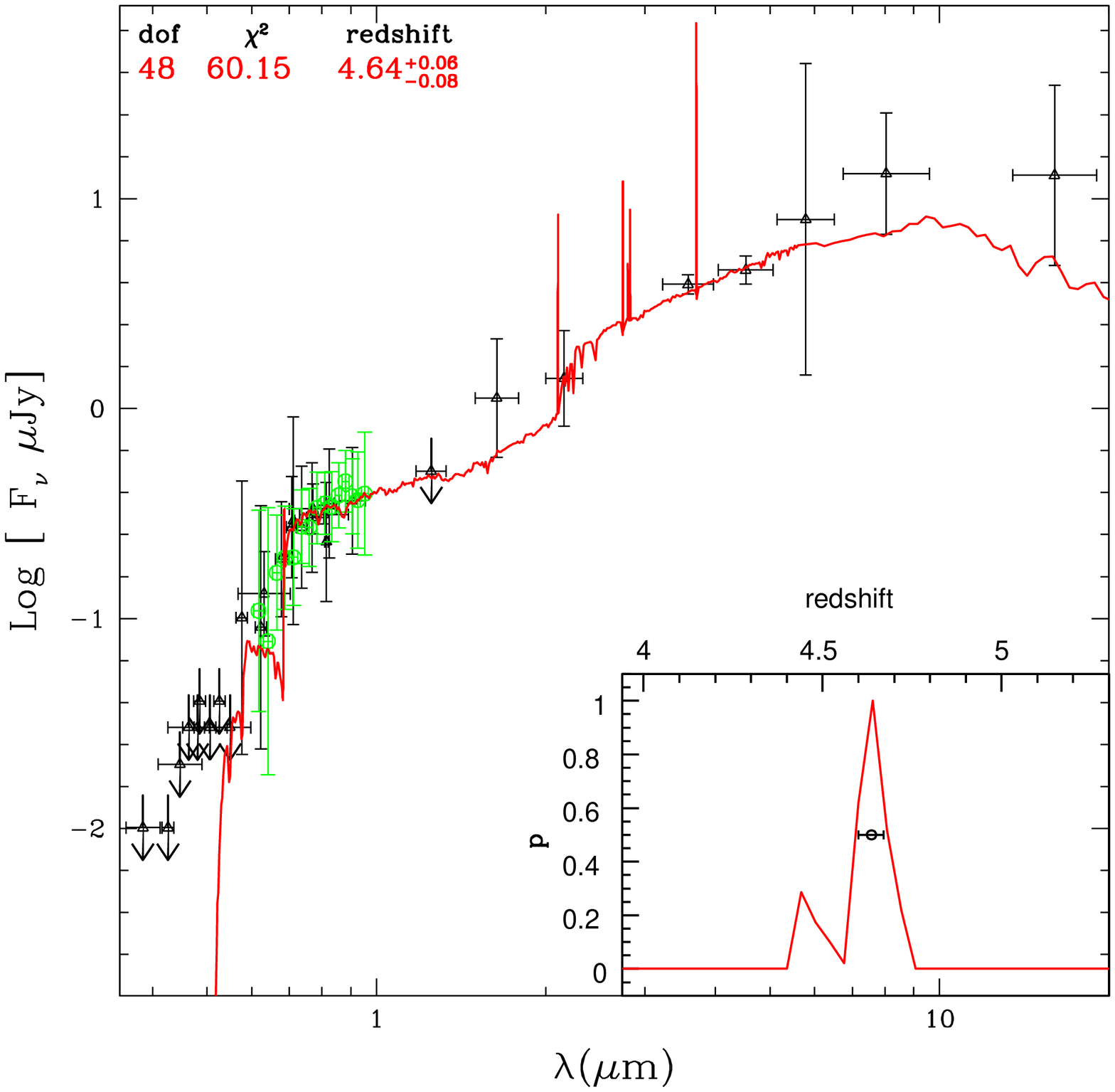} 
\caption{ The UV-IR spectral energy distribution (SED) of AzTEC~1
  (symbols). The spectral template, best fit to the multi-band
    photometry (filled symbols) and the binned DEIMOS spectrum (open
    symbols), redshifted to the most probable redshift ($z=4.64$) is
    also plotted (in red). The redshift probability distribution
  $p\propto\exp{(-0.5\chi^2)}$ is shown in the inset. The median
  redshift and $1\sigma$ uncertainties ($z=4.64^{+0.06}_{-0.08}$), as
  well as the degrees of freedom (dof) and the total $\chi^2$ of the
  best fit are indicated in the top-left.
  \label{fig:specbinn}}
}
\end{figure}

\section{Spectral energy distribution of AzTEC~1}
\label{sec:sed}

In \f{fig:sed} \ we show the SED of AzTEC~1.   Fixing the redshift
  to $z=4.64$ (\s{sec:redsft} ) we fit the UV-NIR SED using various
  model spectrum libraries. For each model we compute the total
  $\chi^2$ and define the most probable parameter values and their
  errors from the probability distribution function. Using the
  \citet{bc03} library (see \citealt{smo08} for details) the UV-NIR
  SED is best described by a $740^{+200}_{-60}$~Myr old starburst with
  $\mathrm{SFR}=410\pm50~\mathrm{M_\odot/yr}$, an extinction of
  $\mathrm{A_V}=2\pm0.2$~mag, and a stellar mass of
  $M_*=(1.5\pm0.2)\times10^{11}~\mathrm{M_\odot}$ (see top panel in
  \f{fig:sed} ). We find consistent results when using the Maraston
  (2005) library.  However, as pointed out by Maraston et al.\
(2010), using exponentially decaying star formation histories as above
some of the free parameters may be poorly parameterized in young
starburst galaxies whose SED is dominated by the youngest stellar
populations that outshine the old ones. Thus, we additionally fit to
the optical-NIR SED of AzTEC~1 the model library presented in
Efstathiou et al.\ (2000), specifically developed for starburst
galaxies. These (UV-mm) models are treated as an ensemble of optically
thick giant molecular clouds (GMCs) centrally illuminated by recently
formed stars. The evolution of the stellar population within the GMC
is modeled using the  \citet{bc03} stellar population synthesis
models. The Efstathiou et al.\ (2000) models yield a
  $37\pm4$~Myr old starburst with $\mathrm{A_V}=100\pm20$ and
  $\mathrm{SFR}=1300\pm150~\mathrm{M_\odot/yr}$. 

We fit the IR portion of the SED of AzTEC~1 (fixing $z=4.64$) using
the Chary \& Elbaz (2001; CE hereafter), Dale \& Helou (2002), and
Lagache et al.\ (2003) models.  The best fit IR model, shown in
\f{fig:sed} \ (bottom panel), is a Lagache et al.\ (2003) template with a total IR
($8-1000~\mathrm{\mu}$m) luminosity of $2.9\times10^{13}$~\lsol , and
a FIR ($60-1000~\mathrm{\mu}$m) luminosity of $9\times10^{12}$~\lsol
. For comparison, the CE SED models yield the second best fit with
integrated luminosities a factor of 3-4 higher. Converting the
($8-1000~\micron$) IR luminosity to a SFR, using
the Kennicutt (1998) conversion, we find a SFR of
$\sim1600~\mathrm{M_\odot/yr}$.  To obtain the dust temperature and
dust mass in AzTEC~1 we perform a gray-body dust model fit to the data
as described in detail in \citet{aravena08}. Using $\beta=1.5$ and
$\beta=2$ we consistently find a dust temperature of
$T_\mathrm{D}\sim50$~K and dust mass of
$M_\mathrm{D}\sim1.5\times10^9$~\msol \ (while the IR luminosity is
within a factor of two compared to that given above).

\begin{figure}[ht!]
\center{\vspace{8mm}
\includegraphics[width=\columnwidth, bb=74 400 466 702]{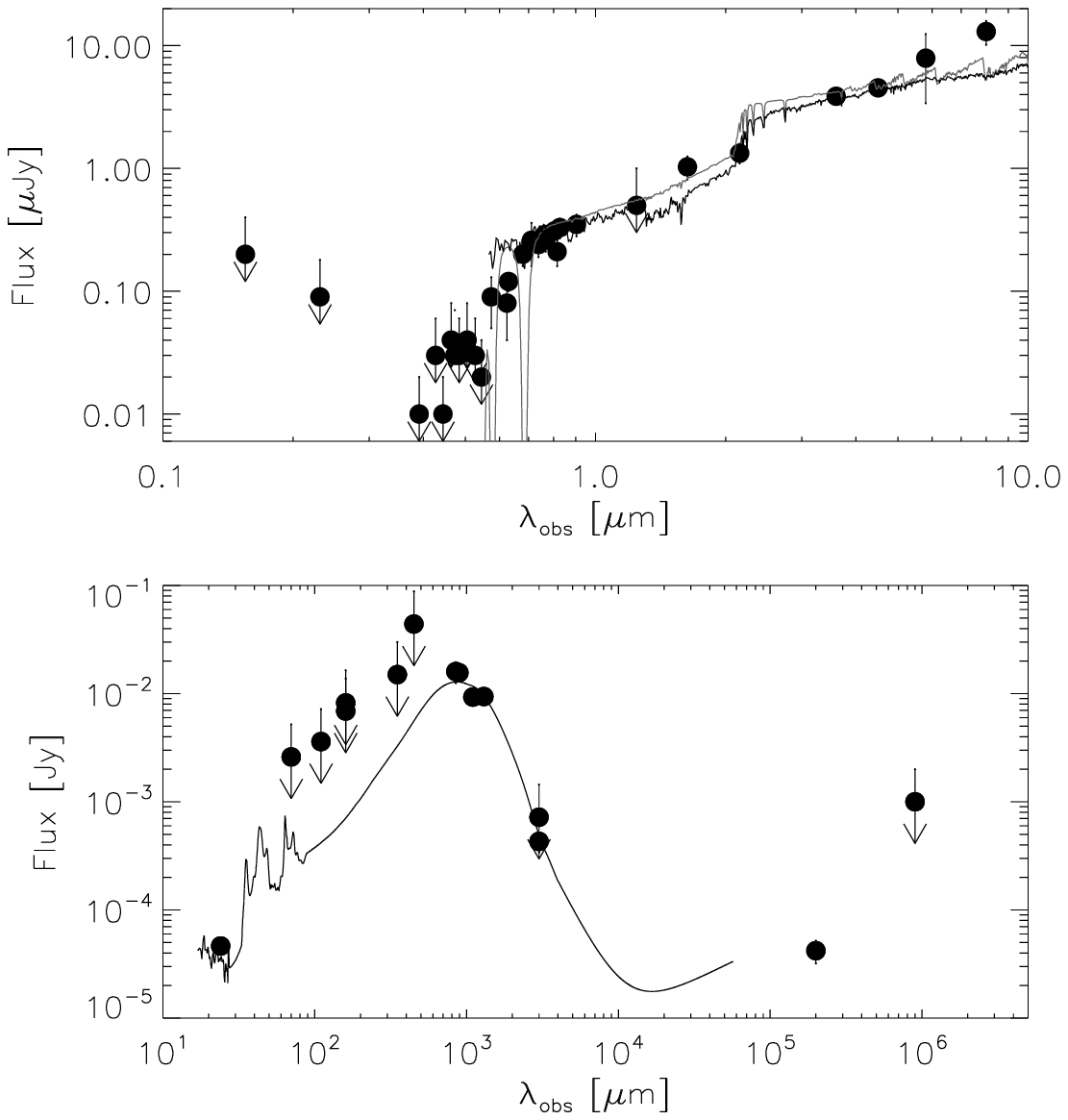}
\caption{ UV-NIR (top) and IR (bottom) SED of AzTEC~1. The best fit
  model spectra from the \citet[][gray]{maraston05} and \citet[][black]{bc03} library to the
  UV-NIR SED and Lagache et al.\ (2003) library to the IR SED are also
  shown (see text for details). 
  \label{fig:sed}}
}
\end{figure}

\section{Discussion}
\label{sec:discussion}

\subsection{Lack of molecular gas?}

Based on observations of AzTEC~1 from radio to X-rays, and a
Keck~II/DEIMOS spectrum, we have shown that AzTEC~1 is a
$L_\mathrm{FIR}=9\times10^{12}$~\lsol\ starburst galaxy at
$z=4.64_{-0.08}^{+0.06}$ (the given errors are $1\sigma$
uncertainties).  However, contrary to expectations our searches for
the CO(5$\rightarrow$4) transition line
($\nu_\mathrm{RF}=576.268$~GHz) in this galaxy with the PdBI/CARMA
interferometers have yielded no detection.  Assuming a line width of
$500$~\kms\ the $3\sigma$ limits in the line luminosity based on PdBI
and CARMA observations are estimated to be
$L'_\mathrm{CO}\lesssim9.8\times10^{9}$~K~\kms ~pc$^2$ ($4.56<z<4.76$)
and $L'_\mathrm{CO}\lesssim6.5\times10^{10}$~K~\kms ~pc$^2$
($4.64<z<4.72$ and $4.94<z<5.02$), respectively.  Taking
$M_\mathrm{gas}/L'_\mathrm{CO}=0.8$~\msol~(K \kms~pc$^2$)$^{-1}$ (Downes \&
Solomon 1998) implies $3\sigma$ gas mass upper limits of
$M_\mathrm{gas}\lesssim8\times10^9$~\msol~($4.56<z<4.76$) and
$M_\mathrm{gas}\lesssim5\times10^{10}$~\msol~$ (4.94<z<5.02$).
Turning the arguments around, i) assuming a typical
$L'_\mathrm{CO}-L_\mathrm{FIR}$ conversion (Riechers et al.\ 2006) the
FIR luminosity inferred here for AzTEC~1,
$L_\mathrm{FIR}=9\times10^{12}$~\lsol , yields an expected CO
luminosity of $L'_\mathrm{CO}\approx4\times10^{10}$~K~\kms ~pc$^2$,
and ii) assuming a gas-to-dust-ratio of 50-150 (e.g.\ Calzetti et al.\
2000), and $M_\mathrm{gas}/L'_\mathrm{CO}=0.8$~\msol~(K \kms~pc$^2$)$^{-1}$ the dust mass we
inferred here for AzTEC~1 ($M_\mathrm{D}\sim1.5\times10^9$~\msol )
translates into a line luminosity of $L'_\mathrm{CO} \sim
(9-30)\times10^{10}$~K~\kms ~pc$^2$. Such a gas reservoir should have
been detected (especially with the more sensitive PdBI observations)
within our interferometric observations in the 3~mm band. Below we
discuss a few possibilities why the CO(5$\rightarrow$4) line was not
detected.

First, it is possible that the systemic redshift of the source is
outside the bandwidth range covered with our interferometric
observations (encompassing redshift ranges of $4.56-4.76$ and
$4.94-5.02$). Our UV-NIR analysis of the SED yields a 68\% probability
that the redshift of the source is within $4.56<z<4.70$. However, we
also find a second redshift peak at $z\sim4.44$ in our redshift
probability distribution (see \f{fig:specbinn} ). Furthermore, the
systemic (CO) redshift of the source is not necessarily expected to
coincide with the one inferred from UV-NIR data (typical velocity
offsets are several hundred \kms\ for narrow-line objects). Thus, it
is possible that the CO redshift is outside the range covered by our
interferometric observations. Note
however, that if this were the case, it would not significantly alter
the results of our SED analysis (\s{sec:sed} ).  Alternatively,
assuming the systemic redshift is within the covered bandwidth, the
CO(5$\rightarrow$4) non-detection could be explained by a low
CO-excitation resulting in a low line brightness of the
CO(5$\rightarrow$4) transition. Assuming a CO 5$\rightarrow$4 to
1$\rightarrow$0 line brightness temperature ratio of $\sim1/3$, as
found for the $z>4$ SMG GN20
\citep{carilli10}, the PdBI $3\sigma$ limit in
  the CO(1$\rightarrow$0) line is
  $L'_\mathrm{CO}\lesssim3\times10^{10}$~K~\kms ~pc$^2$. This is
  roughly consistent with the CO-FIR relation. Furthermore, an
  uncertainty of a factor of a few in the inferred dust mass
  (including possible AGN heating) and the
  $M_\mathrm{gas}/L'_\mathrm{CO}$ conversion factor makes this limit
  also roughly consistent with $L'_\mathrm{CO}$ estimated from
  AzTEC~1's dust mass.  Thus, a low CO-excitation in AzTEC~1 may
  explain the non-detection of CO(5$\rightarrow$4).

\subsection{Mode of star formation}

Our analysis of the UV-radio SED of AzTEC~1 implies that AzTEC~1
is an extremely young and massive galaxy, forming stars at a rate of
$\sim1300$~\msolyr\ at $z=4.6$.  In general, vigorous star formation
induces strong negative feedback that can terminate (and then
self-regulate) the starburst by dispersing and expelling gas from the
gravitational potential well (Elmegreen 1999; Scoville 2003; Thompson
et al.\ 2005; Riechers et al.\ 2009). This sets a number of
physical limits on the starburst. Assuming that a) the maximum
intensity of a radiation-pressure supported starburst is determined by
the Eddington limit for dust, b) a constant gas-to-dust ratio with
radius, and c) that the disk is self-regulated (i.e.\ Toomre
$\mathrm{Q}\sim1$) such an Eddington limited starburst will have a SFR
surface density $\Sigma_\mathrm{SFR}\sim1000$~\msolyrkpc , a FIR
luminosity surface density
$\mathrm{F_{FIR}}\sim10^{13}$~\lsol~kpc$^{-2}$, and an effective
temperature of 88~K (see eqs.\ 33-36 in Thompson et al.\ 2005).

A SFR of $\sim1300$~\msolyr\ in AzTEC~1 (based on the NUV-NIR SED fit)
then implies a SFR surface density of
$\Sigma_\mathrm{SFR}=\mathrm{SFR}/(\pi r^2)\gtrsim420$~\msolyrkpc\
(assuming $r\lesssim1$~kpc based on SMA imaging; Younger et al.\
2008). The inferred value does not violate the Eddington limited
starburst models. The FIR luminosity surface density in AzTEC~1,
$\mathrm{F_{FIR}}=L_\mathrm{FIR}/(\pi
r^2)\gtrsim2.8\times10^{12}$~\lsol~kpc$^{-1}$, and the dust
temperature of $\sim 50$~K support that the starburst in AzTEC~1 is
consistent, but not in violation of its Eddington limit.

It is noteworthy that the inferred value of the SFR surface density
for AzTEC~1 is somewhat higher compared to SMGs at $z\sim2$, which
typically have $\Sigma_\mathrm{SFR}\sim 80$~\msolyrkpc \ (Tacconi et
al.\ 2006), pointing to the compactness of the star formation region in
AzTEC~1. Tacconi~et~al.\ have shown that $z\sim2$ SMGs are well described
within a starburst picture (Elmegreen 1999) in which star formation
cannot self-regulate and thus a significant fraction of gas is
converted into stars in only a few times the dynamical timescale.
Continuing this line of reasoning we make use of a detailed
hydrodynamical study of matter deposition in young assembling galaxies
performed by Silich et al.\ (2010). We estimate that AzTEC~1 is
forming stars in a 'gravitationally bound' regime in which gravity
prohibits the formation of a superwind, leading to matter accumulation
within the galaxy and further generations of star formation.
Specifically, Silich et al.\
show that there are three hydrodynamic regimes that develop in
starbursting galaxies: i)
generation of a
superwind, that expels matter from the star forming region, ii) a
'gravitationally bound' regime, in which gravity prohibits the
formation of a superwind and contains the matter within the galaxy,
and iii) an
intermediate, bimodal regime. The
specific regime is dependent on the SFR and the size of the star formation region
in the galaxy (see Fig.~1 in Silich et al.\ 2010). Taking the size of
the star forming region in AzTEC~1 to be $\sim1$~kpc \citep{younger08}, its
$\mathrm{SFR}\sim1300$~\msolyr\ yields that, consistent
with SCUBA detected galaxies, AzTEC~1 is forming stars in the
gravitationally bound regime.

In summary, our analysis of the properties of AzTEC~1 points to an
extremely young and massive galaxy, forming stars at a rate of
$\sim1300$~\msolyr\ at $z=4.6$.  We find that it has already assembled
a stellar mass of $1.5\times10^{11}$~\msol , in a region covering only
$\sim1-2$~kpc in total extent (based on HST and SMA imaging; see
Younger et al.\ 2007, 2008) yielding that AzTEC~1 is a compact massive
galaxy at $z=4.6$. 

The high stellar mass and compactness of AzTEC~1 resemble that of a recently identified population of quiescent,
passively evolving, already massive (typically
$M_*=1.7\times10^{11}$~\msol ), but compact galaxies at $z\sim2$
(e.g.\ van Dokkum et al.\ 2008) deemed to evolve into the most massive
red-and-dead galaxies at $z\sim0$. The upper gas mass limit inferred
for AzTEC~1 (although quite uncertain) is $\sim10^{10}$~\msol . If AzTEC~1 continues to form stars at the current rate it will deplete
the available gas in $M_\mathrm{gas}/\mathrm{SFR}\sim6$~Myr (assuming 100\%
efficiency). Unless further gas is supplied and high
levels of star formation are induced the galaxy's
stellar body will have time to age and redden  till $z\sim2-3$.

The surface density of the (likely still incomplete) sample of three
confirmed $z>4$ SMGs in the AzTEC-COSMOS field (0.3~deg$^2$) is
$\gtrsim10$~deg$^{-2}$. This is already higher than $\sim7$~deg$^{-2}$
predicted by semi-analytic models of structure growth (e.g. Baugh et
al.\ 2005; see also Coppin et al.\ 2009, 2010). Thus, further studies
of $z>4$ SMGs are key to understand the SMG population (e.g. Wall et
al.\ 2008) and its cosmological role.

\section{Conclusions}

Based on UV-FIR observations of AzTEC~1, and a Keck~II/DEIMOS spectrum, we have shown that AzTEC~1 is a
$L_\mathrm{FIR}=9\times10^{12}$~\lsol\ starburst at $z=4.64_{-0.08}^{+0.06}$ (with a
secondary, less likely, redshift probability peak at $z\sim4.44$). 
Based on our revised FIR values we find that AzTEC~1 fits comfortably
within the limits of a maximal starburst, and that it forms stars in a
gravitationally bound regime which traps the gas within the galaxy
leading to formation of new generations of stars. Our SED analysis
yields that AzTEC~1 is an extremely young ($\lesssim50$~Myr), massive
($M_*\sim10^{11}$~\msol), but compact ($\lesssim2$~kpc) galaxy
forming stars at a rate of $\sim1300$~\msolyr\ at $z=4.64$. These
interesting properties suggest that AzTEC~1 may be a candidate of
progenitors of quiescent, already massive, but very compact galaxies
regularly found at $z\sim2$, and thought to evolve into the most
massive, red-and-dead galaxies found in the local universe.

\acknowledgements The authors acknowledge the
significant cultural role that the summit of Mauna Kea
has within the indigenous Hawaiian community; NASA grants HST-GO-09822 (contracts 1407, 1278386;
SSC); HST-HF-51235.01 (contract NAS 5-26555; STScI); GO7-8136A;
Blancheflor Boncompagni Ludovisi foundation (F.C.); French Agene
National de la Recheche fund ANR-07-BLAN-0228; CNES; Programme
National Cosmologie et Galaxies; UKF; DFG; DFG Leibniz Prize (FKZ HA
1850/28-1); European Union's Seventh Framework programme (grant
agreement 229517); making use of the NASA/ IPAC IRSA,
by JPL/Caltech, under contract with the National Aeronautics and Space
Administration; IRAM PdBI supported by INSU/CNRS (France), MPG
(Germany) and IGN (Spain); CARMA supported by the states of
California, Illinois, and Maryland, the Gordon and Betty Moore
Foundation, the Eileen and Kenneth Norris Foundation, the Caltech
Associates, and NSF.

{}

\begin{table*}
\begin{center}
\caption{AzTEC~1 photometry}
\label{tab:phot}
\vskip 10pt
\footnotesize{
\begin{tabular}{c|c|c}
\hline
    wavelength & band/telescope & flux density ($\mu$Jy) \\
\hline
0.5-2~keV       & Chandra-soft-band & $<0.0003^{a,b}$ \\
1551\AA 	& FUV   & $<0.20^a$  \\
2307\AA 	& NUV   & $<0.09^a$  \\
3911\AA 	& u$^*$ & $<0.01^a$  \\
4270\AA 	& IA427 & $<0.03^a$  \\
4440\AA 	& B$_J$ & $<0.01^a$  \\
4640\AA 	& IA464 & $<0.04^a$  \\
4728\AA 	& g$^+$ & $<0.03^a$  \\
4840\AA 	& IA484 & $<0.03^a$  \\
5050\AA 	& IA505 & $<0.04^a$  \\
5270\AA 	& IA527 & $<0.03^a$  \\
5449\AA 	& V$_J$ & $<0.02^a$ \\
5740\AA		& IA574	& 0.09	$\pm$ 0.04\\
6240\AA		& IA624	& 0.08	$\pm$ 0.04\\
6295\AA		& r$^+$	& 0.12	$\pm$ 0.02\\
6790\AA		& IA679	& 0.20	$\pm$ 0.04\\
7090\AA		& IA709	& 0.25	$\pm$ 0.04\\
7110\AA		& NB711	& 0.26	$\pm$ 0.10\\
7380\AA		& IA738	& 0.24	$\pm$ 0.05\\
7641\AA		& i$^+$	& 0.29	$\pm$ 0.02\\
7670\AA		& IA767	& 0.26	$\pm$ 0.05\\
8040\AA		& F814W	& 0.31	$\pm$ 0.02\\
8150\AA		& NB816	& 0.21	$\pm$ 0.05\\
8270\AA		& IA827	& 0.33	$\pm$ 0.06\\
9037\AA		& z$^+$	& 0.35	$\pm$ 0.07\\
12444\AA        & J     & $<0.5^a$\\
16310\AA	& H	& 1.03	$\pm$ 0.22\\
21537\AA	& K$_s$	& 1.33	$\pm$ 0.23\\
3.6~$\mu$m	& IRAC1	& 3.87	$\pm$ 0.13\\
4.5~$\mu$m	& IRAC2	& 4.53	$\pm$ 0.23\\
5.8~$\mu$m	& IRAC3	& 7.90	$\pm$ 4.50\\
8.0~$\mu$m	& IRAC4	& 13.01	$\pm$ 2.88\\
16~$\mu$m	& IRS-16  & 12.80 $\pm$ 4.20\\
24~$\mu$m	& MIPS-24 & 46.40 $\pm$ 4.90\\
70~$\mu$m & MIPS-70 & $<2600^a$ \\
100~$\mu$m & PACS-100 & $<3600^a$ \\
160~$\mu$m & MIPS-160 & $<8200^a$ \\
160~$\mu$m & PACS-160 & $<6900^a$ \\
350~$\mu$m & CSO & $<15000^a$ \\
450~$\mu$m & SCUBA-2 & $<44000^a$ \\
850~$\mu$m & SCUBA-2 & $16000\pm3500$ \\
890~$\mu$m & SMA & 15600 $\pm$ 1100 \\
1.1~mm & JCMT/AzTEC & 9300 $\pm$ 1300 \\
1.3~mm & CARMA & 9400 $\pm$ 1600 \\
3~mm & CARMA &  $<720^a$\\
3~mm & PdBI & $300\pm40$ \\
20~cm & VLA & $42.0\pm10$ \\
90~cm & VLA & $<1000^a$
\end{tabular}
}
\end{center}
$^a$~The given limits are $2\sigma$ upper limits.
$^b$~Corresponds to $=10^{-15}$~erg~cm$^{-2}$~s$^{-1}$ 
\end{table*}

\end{document}